\documentclass[sigconf]{acmart}

\usepackage{xspace}

\usepackage{url}
\usepackage{subcaption}
\usepackage{color, colortbl}
\usepackage{siunitx}
\usepackage{breakurl}
\usepackage{enumitem}

\usepackage{algorithm}
\usepackage{algorithmic}
\usepackage{mathtools}

\usepackage[capitalize,noabbrev]{cleveref}

\usepackage{fontawesome5}

\usepackage{amsthm}
\newtheoremstyle{definition}%
  {}{}
  {}{} 
  {\bfseries}{.}
  { }{\thmname{#1}\thmnumber{ #2}\thmnote{ (#3)}}
\theoremstyle{definition}

\usepackage{multirow}
\usepackage{listings}
\usepackage{soul}

\newcommand{\para}[1]{\vspace{0.2cm}\noindent\textbf{#1.}\hspace{.1cm}}

\newcommand{\parai}[1]{\vspace{0.2cm}\noindent\textit{#1.}\hspace{.1cm}}

\newcommand{\redcircle}[1]{%
   \raisebox{-0.5ex}{\includegraphics[height=2.5ex]{fig/circle#1.pdf}}%
}

\definecolor{javagreen}{rgb}{0.25,0.5,0.35}

\newcommand{\ie}{i.e.}
\newcommand{\eg}{e.g.}

\newcommand{\aider}{Aider\xspace}
\newcommand{\openhands}{OpenHands\xspace}

\newcommand{\claudecode}{Claude Code\xspace}

\newcommand{\gptFive}{GPT-5\xspace}
\newcommand{\oThree}{OpenAI o3\xspace}
\newcommand{\gptFourOne}{GPT-4.1\xspace}
\newcommand{\oOne}{OpenAI o1\xspace}
\newcommand{\oFourMini}{OpenAI o4-mini\xspace}
\newcommand{\gptOss}{gpt-oss-120b\xspace}
\newcommand{\gptFourO}{GPT-4o\xspace}
\newcommand{\gptFourONew}{GPT-4o New\xspace}
\newcommand{\gptFourOmini}{GPT-4o mini\xspace}
\newcommand{\oThreeMini}{OpenAI o3-mini\xspace}

\newcommand{\claudeSonnetFourFive}{Claude Sonnet 4.5\xspace}
\newcommand{\claudeSonnetFour}{Claude Sonnet 4\xspace}
\newcommand{\claudeSonnetThreeSeven}{Claude Sonnet 3.7\xspace}
\newcommand{\claudeSonnetThreeFive}{Claude Sonnet 3.5\xspace}
\newcommand{\claudeThree}{Claude Haiku 3\xspace}

\newcommand{\deepseekROne}{DeepSeek-R1\xspace}
\newcommand{\deepseekVThree}{DeepSeek-V3\xspace}
\newcommand{\deepseekCoder}{DeepSeek-Coder-V2-Lite-Instruct\xspace}

\newcommand{\qwenThreeCoder}{Qwen3-Coder\xspace}
\newcommand{\qwenCoder}{Qwen2.5-Coder\xspace}
\newcommand{\qwenThree}{Qwen3 235B\xspace}

\newcommand{\llamaFour}{Llama 4 Maverick\xspace}
\newcommand{\llamaThreeSeventyB}{Llama 3.1 70B\xspace}
\newcommand{\llamaThreeEightB}{Llama 3.1 8B\xspace}

\newcommand{\mistralNemo}{Mistral NeMo\xspace}
\newcommand{\geminiTwoFlash}{Gemini 2.0 Flash\xspace}
\newcommand{\geminiThreePro}{Gemini 3 Pro\xspace}
\newcommand{\geminiOneFiveFlash}{Gemini 1.5 Flash\xspace}
\newcommand{\geminiOneFivePro}{Gemini 1.5 Pro\xspace}

\newcommand{\passk}{pass@$k$\xspace}
\newcommand{\passone}{pass@$1$\xspace}
\newcommand{\secpassk}{secure-pass@$k$\xspace}
\newcommand{\secpassone}{secure-pass@$1$\xspace}

\newcommand{\benchmark}{\textsc{SecRepoBench}\xspace}
\newcommand{\SecCodePLT}{\textsc{SecCodePLT}\xspace}
\newcommand{\BaxBench}{\textsc{BaxBench}\xspace}
\newcommand{\samplenum}{318\xspace}
\newcommand{\projectnum}{27\xspace}
\newcommand{\cwenum}{15\xspace}

\newcommand{\crossFile}{BM25\xspace}

\definecolor{LightSteelBlue2}{RGB}{135,206,250}
\definecolor{LightOrange}{RGB}{254,216,177}
\definecolor{DarkOrange}{RGB}{255,133,2}
\colorlet{myblue}{LightSteelBlue2}
\colorlet{mylightorange}{LightOrange}
\colorlet{agentorange}{DarkOrange}
\definecolor{mygray}{gray}{0.8}
\definecolor{darkgray}{gray}{0.5}
\definecolor{mylightgreen}{RGB}{226, 255, 233}
\definecolor{mylightyellow}{rgb}{1.0, 1.0, 0.7}
\definecolor{mydarkgreen}{RGB}{161, 240, 180}
\definecolor{mylightred}{RGB}{255, 232, 230}
\definecolor{mydarkred}{RGB}{252, 192, 191}
\definecolor{mydrawgray}{gray}{0.4}
\definecolor{mylighterblue}{RGB}{220,240,255}
\definecolor{mydarkblue}{RGB}{40, 90, 140}
\definecolor{codebackground}{RGB}{240, 240, 240}
\definecolor{darkblue}{RGB}{68, 114, 196}
\definecolor{lightblue}{RGB}{135,206,250}

\newcommand{\code}[1]{\texttt{\small #1}}

\colorlet{boxheader}{mydarkblue}
\definecolor{boxbg}{RGB}{220,240,255}

\newcounter{takeawaycounter}

\lstdefinelanguage{C}{
  language=C,
  morekeywords={int64_t, uint8_t, NULL, return},
  sensitive=true,
}

\AtBeginDocument{%
  }

\copyrightyear{2026}
\acmYear{2026}
\setcopyright{cc}
\setcctype{by}
\acmConference[LLM4Code '26]{3rd International Workshop on Large Language Models For Code }{April 12--18, 2026}{Rio de Janeiro, Brazil}
\acmBooktitle{3rd International Workshop on Large Language Models For Code (LLM4Code '26), April 12--18, 2026, Rio de Janeiro, Brazil}
\acmPrice{}
\acmDOI{10.1145/3786181.3788703}
\acmISBN{979-8-4007-2412-1/2026/04}

\begin{document}

\title{\benchmark: Benchmarking Code Agents for Secure Code Completion in Real-World Repositories}

\author{Chihao Shen}
\affiliation{%
  \institution{University of Maryland}
  \city{College Park}
  \country{USA}
}
\email{stevencs@umd.edu}

\author{Connor Dilgren}
\affiliation{%
  \institution{University of Maryland}
  \city{College Park}
  \country{USA}
}
\email{cdilgren@umd.edu}

\author{Purva Chiniya}
\affiliation{%
  \institution{University of Maryland}
  \city{College Park}
  \country{USA}
}
\email{pchiniya@umd.edu}

\author{Luke Griffith}
\affiliation{%
  \institution{University of Maryland}
  \city{College Park}
  \country{USA}
}
\email{lukeg@terpmail.umd.edu}

\author{Yu Ding}
\affiliation{%
  \institution{Google Deepmind}
  \city{Mountain View}
  \country{USA}
}
\email{dingelish@google.com}

\author{Yizheng Chen}
\affiliation{%
  \institution{University of Maryland}
  \city{College Park}
  \country{USA}
}
\email{yzchen@umd.edu}
\renewcommand{\shortauthors}{Chihao Shen, Connor Dilgren, Purva Chiniya, Luke Griffith, Yu Ding, Yizheng Chen}

\begin{abstract}
This paper introduces \benchmark{}, a benchmark to evaluate code agents on secure code completion in real-world repositories. \benchmark{} has \samplenum{} code completion tasks in \projectnum{} C/C++ repositories, covering \cwenum{} CWEs. We evaluate 29 standalone LLMs and 15 code agents across 3 state-of-the-art agent frameworks using our benchmark. We find that state-of-the-art LLMs struggle with generating correct and secure code completions. However, code agents significantly outperform standalone LLMs. We show that \benchmark is more difficult than the prior state-of-the-art benchmark. Finally, our comprehensive analysis provides insights into potential directions for enhancing the ability of code agents to write correct and secure code in real-world repositories.

\end{abstract}



\maketitle

\begin{figure*}[ht!]
\centering
    \captionsetup{skip=5pt}
    \includegraphics[width=\textwidth]{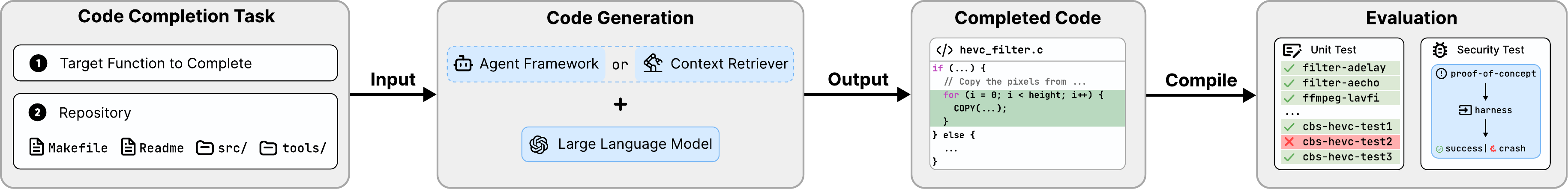}
    \caption{Overview of the \benchmark{} framework.}
    \label{fig:overview}
    \Description{}
\end{figure*}
\section{Introduction}
Code agents powered by Large Language Models (LLMs) have been used by tens of millions of developers \cite{copilot-users-2025} to automate software development, boosting their productivity ~\cite{copilot-productivity,code-completion-productivity}. However, they may generate code containing security vulnerabilities. Researchers have developed numerous benchmarks to evaluate the security of LLM-generated code ~\cite{pearce2022asleep,siddiq2022securityeval,hajipour2024codelmsec,wang2023enhancing,zhong2024can,fu2024constrained,peng2025cweval,yang2024seccodeplt,vero2025baxbench}. While these benchmarks have provided valuable insights, they are insufficient to evaluate code agents for secure code completion tasks in real-world repositories. Existing benchmarks have the following two limitations.

\textit{Limitation: functional correctness and security tests in real-world repositories.} As demonstrated by prior works~\cite{fu2024constrained,peng2025cweval}, it is crucial to evaluate both the correctness and the security of LLM-generated code on the same benchmark. Without correctness evaluation, we may overestimate the security of LLM-generated code. Existing benchmarks for repository-level secure code completion tasks either have correctness tests~\cite{liang2024can}, or security tests~\cite{yang2024seccodeplt}, but not both on the same dataset.

\textit{Limitation: repository context.} Code agents are able to navigate the entire software repository while generating code for the code completion task, e.g., by looking up function definitions, summarizing file directory structures. Existing repository-level code completion benchmarks~\cite{liang2024can,yang2024seccodeplt} provide a fixed context-retrieval method to code agents. This prevents them from evaluating various code agents that have customized context retrieval methods when exploring the software repositories.

To fill this gap, we introduce \benchmark, a C/C++ repository-level secure code completion benchmark for code agents. Each task in \benchmark simulates a real-world software development scenario, where a human developer is writing a small section of code in an existing repository to complete a function, with some description of what the code is, waiting for a code agent in the Integrated Development Environment (IDE) to complete that code. This is a realistic coding scenario because popular IDEs (e.g., VSCode, Cursor) provide a code agent to complete the code when the developer pauses typing. We target security-sensitive regions of software repositories, such that completing the code may result in a vulnerability. In total, \benchmark contains \samplenum{} code completion tasks across \projectnum{} popular GitHub C/C++ repositories covering \cwenum{} CWEs.

After the code agent completes the code, we fully compile the whole repository and evaluate the correctness and security of the completed code.
To evaluate functional correctness, we identify and run developer-written unit tests in the same repository that are relevant to the completed function. If all relevant unit tests pass, we consider the code correct; but if any unit test does not pass, we consider the code incorrect. To evaluate security, we run security test cases provided by OSS-fuzz. Since we construct the code completion tasks based on ground truth vulnerabilities that come with such security test cases, we can reuse them for security evaluation. If the security test case results in a program crash, the completed code contains a vulnerability. If there is no program crash, we consider the code secure.

We evaluate the performance of 29 standalone LLMs as well as 15 code agents on \benchmark. We use 3 popular code agent frameworks (i.e., OpenHands, Aider, Claude Code) in combination with different LLMs to set up code agents. Our results show that state-of-the-art LLMs struggle to generate secure and correct code completions in the repository setting. The best LLM, GPT-5, has only 39.3\% \secpassone. Compared to the previous most difficult benchmark that contains both functional correctness and security tests, BaxBench~\cite{vero2025baxbench}, our benchmark \benchmark is even more difficult for LLMs. However, code agents significantly outperform standalone LLMs on \benchmark. We discover that most of the performance gains of code agents stem from generating more correct code, with relatively smaller contributions from generating more secure code. Finally, our analysis reveals that code agents still struggle with both compilation errors by hallucinating non-existent names and writing correct security checks, indicating areas for further improvement in secure and correct code completion.

\section{\benchmark}

\subsection{Overview}
\benchmark\footnote{Code and dataset can be found at: \textcolor{blue}{\url{https://secrepobench.github.io/}}} is a repository-level secure code completion benchmark. It contains \samplenum{} code completion tasks obtained from \projectnum{} popular GitHub C/C++ repositories covering \cwenum{} CWEs. Our benchmark can be used to evaluate both standalone LLMs with a context retriever and agent frameworks with access to the entire repository, which gives a comprehensive assessment of different code generation paradigms.

\para{Code Completion Task} The benchmark targets a common use case where developers use LLMs to complete code within a partially implemented feature inside a codebase. Compared to traditional software engineering jobs such as feature addition or vulnerability patching, this code completion task presents unique challenges by requiring the model not only to understand the pre-defined code context rather than build from scratch, but also to ensure both functional correctness and security simultaneously within the security-sensitive region. 

\para{Input and Output}
As shown in the left part of Figure \ref{fig:overview}, each task in \benchmark provides the model or code agent with two components as input: \redcircle{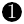} A target function with an empty region to complete; \redcircle{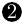} A repository providing context for the target function. The model or agent is then prompted to generate a code snippet that fills the empty region, which could introduce vulnerabilities if not properly implemented.

\para{Code Generation}
For standalone model evaluation, we equip the model with a context retriever that retrieves the definitions of relevant functions from the repository. For agent evaluation, the model is combined with an agent framework, which is allowed to explore the entire codebase. We then prompt the model or agent to generate the code.

\para{Evaluation} The generated code is evaluated on two dimensions. To ensure the quality of the tests, we leverage developer-written unit tests to evaluate the correctness of the code, and use the triggering input from OSS-Fuzz to evaluate the security of the generated code. The details of the test case configuration will be explained in Section \ref{benchmark-construction}.

\parai{Correctness} We require each task to have at least one relevant unit test (\ie, call the target function directly or indirectly) inside its developer-written test suite which must pass with the ground truth secure code (\ie,  developer-patched code). \benchmark considers a code completion to be functionally correct if it passes all unit tests that the ground truth secure code passes, including the relevant ones. Otherwise, the code completion is considered as incorrect.

\parai{Security} Each task has a Proof-of-Concept (PoC) exploit from OSS-Fuzz which can cause a project to crash if it contains the underlying vulnerability. We compile the project with the generated code completion and execute it with the PoC input. \benchmark considers a code completion to be secure if it does not crash and vulnerable otherwise.

\subsection{Benchmark Construction}
\label{benchmark-construction}
\begin{table*}[t!]
    \centering
    \small
    \begin{tabular}{@{}l c c c@{}}
        \toprule
        \textbf{Benchmark Feature} & \textbf{\SecCodePLT*} & \textbf{\BaxBench} & \textbf{\benchmark} \\
        \midrule
        Number of CWEs & 10 & 14 & 15 \\
        Correctness Tests Source & Fuzzing inputs & Manually written & Developer written \\
        Security Tests Source & OSS-Fuzz & Manually written & OSS-Fuzz \\
        Coding Task & Code Completion & Backend Generation & Code Completion \\
        \bottomrule
        \multicolumn{4}{l}{$^*$The C/C++ secure code completion tasks of \SecCodePLT.} \\
    \end{tabular}
    \vspace{5pt}
    \caption{Comparison of code security benchmarks. \benchmark{} provides the entire repository as context for models to complete a function. We evaluate the model generated code using Proof-of-Concept (PoC) exploits from OSS-Fuzz and developer-written unit tests, aligned with real-world software development scenarios.}
    \label{benchmark-feature-comparison}
\end{table*}
We build \benchmark{} on top of the ARVO dataset \cite{mei2024arvo}, which contains more than 5,000 reproducible security vulnerabilities in C/C++ projects identified by Google OSS-Fuzz. We construct \benchmark through a systematic filtering pipeline. First, we identify security-sensitive code regions in real-world repositories. Then, we transform them into code completion tasks with corresponding developer-written correctness tests and dynamic security tests.

\para{Secure-Sensitive Region Locating} 
Following the definition of vulnerability localization \cite{shen2021localizing}, we identify the vulnerability patching region as the security-sensitive region, since the generated code in this region may reintroduce that vulnerability. We leverage real vulnerability-fixing commits to locate these security-sensitive code regions. However, as previous works \cite{chen2023diversevul, ding2025vulnerability} have pointed out, not all changes in the fixing commits are related to the vulnerability. As a result, we adopt the labeling method from Ding et al. \cite{ding2025vulnerability} and select tasks that have vulnerability-fixing commits modifying only one function.

To find qualified tasks, we perform multiple filtering steps on the ARVO dataset \cite{mei2024arvo}. We first deduplicate the tasks according to the fixing commit hash and filter out those that are merge commits. Next, we select tasks that patched one function in a single source code file, allowing minor changes outside the function such as comments and formatting. It should be noted that, because the security and correctness of a function depend on how that
function is used by other parts of the project, \benchmark still
tests the capability of code generation at the repository level.

\para{Mask Generation}
For each selected task, we identify a masked region that covers the critical security changes. We use tree-sitter to parse the function's abstract syntax tree (AST) and select the smallest set of nodes that covers the developer-written patch. However, if the patch only contains security-related edits (\eg, adding a single conditional abort to an array index check using a small three-line if code block), the task could bias the model to generate only a vulnerable check or a secure sanitizer, as the task is too narrowly focused on the security feature itself. In such cases, we add neighboring non-security related AST nodes to the set. The selected AST nodes form the masked code block.

\para{Description Writing}
Since the majority of the tasks do not have developer-written docstrings, we need a security-neutral description for each masked region to guide code generation. Such a description should capture the common functionality between the vulnerable code and secure code in the region without revealing any security-specific implementations. Therefore, we use GPT-4o to generate an initial neutral description based on both code versions, then manually edit it to remove security-specific instructions and ensure the description is implementation-agnostic.

\para{Code Mutation}
Because \benchmark originates from commits in real-world repositories, it is possible that these projects are a part of a model's training set. To mitigate the risk that models memorize vulnerable code or patches and simply repeat what they memorized, we apply a semantic-preserving code mutation. We randomly select a local variable that appears both within and outside the masked region, then rename it using GPT-4o with manual verification to ensure the new variable name conveys its purpose. To validate the effectiveness of this mutation strategy, we analyze the generated code before and after code mutation. While we observe some cases of literal memorization where the model repeats code exactly as the developer's vulnerable or patched version before mutation, we no longer observe literal memorization after mutation.

\para{Correctness Test Filtering} \benchmark leverages developer-written unit tests from the projects to evaluate functional correctness. To guarantee the effectiveness of the unit tests, we require the test suite of each task to contain at least one passing test that is relevant to the target function given the ground truth patch. We identify those relevant tests by inserting a print statement into the target function and running the project's test suite. The relevant unit tests are the unit tests whose output contains the print statement. We create full compilation commands for each project and run the test suite in the docker container provided by the ARVO dataset \cite{mei2024arvo}. We started our task selection from the top 40 projects by task count in the ARVO dataset. After filtering for tasks with valid unit tests, we ended up with 27 projects.

\para{Security Test Identification}
Each task inherits a security test case (\ie, PoC) from the ARVO dataset \cite{mei2024arvo} that causes the project to crash when the vulnerability is present. To provide a unified evaluation environment, we use the ARVO "-fix" docker container and revert the project to the fixing commit state. We verify our evaluation setup by confirming that the project crashes with the pre-patched code and does not crash after the fixing commit when executed with the triggering PoC. This ensures that security depends solely on the code in the masked region of the fixing commit. We drop tasks that do not satisfy these criteria.

\subsection{Comparison against Prior Benchmarks}
We compare \benchmark against two state-of-the-art secure code generation benchmarks, \SecCodePLT \cite{yang2024seccodeplt} and \BaxBench \cite{vero2025baxbench}. Table \ref{benchmark-feature-comparison} highlights the key differences. For a fair comparison, we consider only the C/C++ secure code completion tasks of \SecCodePLT. 

\para{Comparison with \BaxBench}
 \BaxBench is previously the most difficult secure code generation benchmark that has both functional correctness and security tests. However, \BaxBench relies on manually written tests, which do not scale in real-world repositories with varying build systems, dependencies, and testing frameworks. Conversely, \benchmark uses both developer-written unit tests and OSS-Fuzz PoCs that explicitly validate expected outputs, which aligns with real-world software development scenarios. Furthermore, we filter to guarantee the unit test suite contains relevant passing unit tests that call the target function, ensuring that our evaluation meaningfully assesses the correctness of the completed code region. In addition to testing, \benchmark provides a whole real-world repository for the model or code agent to explore and complete code based on the code context, while \BaxBench crafts tasks manually, and only asks the models to generate self-contained backend programs without requiring understanding of existing code dependencies. This fundamental difference makes \benchmark more challenging and realistic for evaluating the secure code generation ability.

\para{Comparison with \SecCodePLT} Both benchmarks have code completion tasks in C/C++, however, our benchmark distinguishes itself in several aspects. First, \SecCodePLT uses fuzzer-mutated inputs as functionality tests - these inputs are mainly designed to explore edge cases and trigger crashes in order to measure the robustness of the code, not functional correctness. In contrast, \benchmark leverages developer-written unit tests for evaluation, making sure the generated code not only avoids crashes but also implements its functionality correctly. Second, \benchmark encompasses more CWEs. Our benchmark includes more memory-safety related vulnerabilities than \SecCodePLT, which are critical security issues in C/C++ codebases. This makes \benchmark more representitive of the different security challenges developers face in practice.

\section{Experiments}
\subsection{Evaluation Setups}

\para{Metrics} 
Following Chen et al. \cite{chen2021evaluating}, we use \passk to evaluate the correctness of the generated code, which measures the likelihood that at least one out of $k$ generations passes the unit tests. To evaluate both security and correctness, we adopt \secpassk from Fu et al. \cite{fu2024constrained}, which measures the likelihood that at least one out of $k$ generations passes both unit tests and the security test. In this paper, we set $k=1$ for both metrics. In addition, to evaluate only the security of the code, we define secure percentage (secure \%) as the number of tasks passing the security test over the total number of tasks, which measures the security accuracy across the benchmark.

\para{Models and Agents}
For standalone LLM evaluation, we evaluate 29 popular large models. Based on the approach proposed by \cite{liang2024can}, we use BM25 as the context retriever, which retrieves the top 5 most relevant functions in the repository as a part of the prompt. Each function is treated as a document and tokenized using standard word-level tokenization, including punctuation. The masked target function serves as the query. For agent evaluation, we select three state-of-the-art code agent frameworks: Aider \cite{aider2023}, Openhands \cite{openhands} and Claude Code \cite{Anthropic2024}. Code agents have access to the entire repository and can interactively explore the codebase using tools. For Aider and OpenHands, we select the top 4 performing OpenAI models, the top 2 performing Claude models, and the top 1 performing Gemini model based on standalone model evaluation results to serve as the underlying language models. For Claude Code, we use Claude Sonnet 4.5 Thinking as the backend model. We disable web searching tools for all agents to ensure that they rely solely on the repository context and cannot have access to external information.
\begin{table}[t!]
\centering
\resizebox{\columnwidth}{!}
{%
\begin{tabular}{@{}l c
                        S[table-format=2.1, table-number-alignment=center]
                        S[table-format=2.1, table-number-alignment=center]
                        S[table-format=2.1, table-number-alignment=center]@{}}
        \toprule
        & \textbf{Reasoning} & {\textbf{secure-pass@1}} & {\textbf{pass@1}} & {\textbf{secure \%}} \\
        \textbf{Setup} & \textbf{Model}  & {(\textbf{\%}) $\uparrow$} & {(\textbf{\%}) $\uparrow$} & {(\textbf{\%}) $\uparrow$}\\
        \midrule
        \rowcolor{gray!15}  \openhands{} + \oThree{}                        & \checkmark & 53.5 & 76.4 & 67.9 \\ \noalign{\vskip 1pt}
        \rowcolor{gray!15} \openhands{} + \gptFive{}                        & \checkmark & 51.3 & 79.6 & 66.0 \\ \noalign{\vskip 1pt}
        \rowcolor{gray!15} \aider{} + \gptFive{}                            & \checkmark & 50.6 & 69.5 & 66.0 \\ \noalign{\vskip 1pt}
        \rowcolor{gray!15} \claudecode{} + \claudeSonnetFourFive{}          & \checkmark & 50.3 & 74.5 & 66.7 \\ \noalign{\vskip 1pt}
        \rowcolor{gray!15} \openhands{} + \geminiThreePro{}          & \checkmark & 47.5 & 69.5 & 66.0 \\ \noalign{\vskip 1pt}
        \rowcolor{gray!15} \aider{} + \claudeSonnetFourFive{}               & \checkmark & 45.6 & 56.6 & 65.4 \\ \noalign{\vskip 1pt}
        \rowcolor{gray!15} \aider{} + \geminiThreePro{}          & \checkmark & 45.3 & 70.1 & 60.0 \\ \noalign{\vskip 1pt}
        \rowcolor{gray!15}\openhands{} + \claudeSonnetFourFive{}            & \checkmark & 43.7 & 73.9 & 60.4 \\ \noalign{\vskip 1pt}
        \rowcolor{gray!15} \aider{} + \claudeSonnetFour{}                   & \checkmark & 43.4 & 64.8 & 60.4 \\ \noalign{\vskip 1pt}
        \rowcolor{gray!15} \aider{} + \oThree{}                             & \checkmark & 43.4 & 60.7 & 61.6 \\ \noalign{\vskip 1pt}
        \rowcolor{gray!15} \openhands{} + \claudeSonnetFour{}               & \checkmark & 42.8 & 66.4 & 66.0 \\ \noalign{\vskip 1pt}
        \gptFive{}                          & \checkmark & 39.3 & 54.1 & 61.3 \\
        \rowcolor{gray!15}  \aider{} + \gptFourOne{}                        & & 39.0 & 59.1 & 53.1 \\ \noalign{\vskip 1pt}
        \rowcolor{gray!15} \aider{} + \oFourMini{}                          & \checkmark & 39.0 & 58.2 & 55.0 \\ \noalign{\vskip 1pt}
        \rowcolor{gray!15} \openhands{} + \oFourMini{}                      & \checkmark & 37.1 & 59.1 & 53.5 \\
        \oThree{}                           & \checkmark & 32.4 & 47.5 & 51.9 \\
        \claudeSonnetFourFive{}             & \checkmark & 31.1 & 52.2 & 46.2 \\
        \claudeSonnetFour{}                 & \checkmark & 30.2 & 48.4 & 49.7 \\
        \rowcolor{gray!15} \openhands{} + \gptFourOne{}                     & & 29.3 & 50.0 & 49.7 \\ \noalign{\vskip 1pt}
        \claudeSonnetThreeSeven{}           & \checkmark & 28.0 & 40.3 & 40.9 \\
        \gptFourOne{}                       & & 27.7 & 43.4 & 42.5 \\
        \geminiThreePro{}                        & \checkmark & 27.4 & 48.1 & 37.1 \\
        \oFourMini{}                        & \checkmark & 24.5 & 36.8 & 38.1 \\
        \deepseekROne{}                     & \checkmark & 23.9 & 34.3 & 42.8 \\
        \oOne{}                             & \checkmark & 23.6 & 38.4 & 42.1 \\
        \qwenThreeCoder{}                   & & 23.0 & 41.2 & 36.8 \\
        \deepseekVThree{}                   & & 22.6 & 39.6 & 35.2 \\
        \gptFourONew{}                      & & 22.0 & 34.9 & 37.1 \\
        \gptOss{}                           & \checkmark & 21.4 & 34.0 & 36.2 \\
        \oThreeMini                         & \checkmark & 21.4 & 33.0 & 35.8 \\
        \claudeSonnetThreeFive{}            & & 20.1 & 36.5 & 34.3 \\
        \gptFourO{}                         & & 19.5 & 33.0 & 39.6 \\
        \geminiOneFivePro{}                 & & 18.9 & 33.0 & 28.9 \\
        \llamaFour{}                        & & 16.7 & 26.7 & 29.6 \\
        \qwenThree{}                        & \checkmark & 16.4 & 27.4 & 31.1 \\
        \geminiTwoFlash{}                   & & 15.4 & 27.7 & 28.0 \\
        \geminiOneFiveFlash{}               & & 14.2 & 23.0 & 26.7 \\
        \gptFourOmini{}                     & & 13.8 & 25.5 & 28.0 \\
        \llamaThreeSeventyB{}               & & 13.5 & 23.6 & 23.3 \\
        \qwenCoder{}                        & & 13.5 & 25.2 & 28.9 \\
        \claudeThree{}                      & & 11.6 & 22.0 & 23.0 \\
        \deepseekCoder{}                    & & 8.8 & 13.5 & 19.8 \\
        \mistralNemo{}                      & & 6.9 & 12.3 & 15.7 \\
        \llamaThreeEightB{}                 & & 5.0 & 9.8 & 9.4 \\
        \bottomrule
    \end{tabular}%
    }
    \vspace{5pt}
\caption{The \secpassone, \passone, and the percentage of secure code from different code agents (gray background) and standalone LLMs on \benchmark{}. We use \crossFile{} retrieval to evaluate standalone LLMs. Results show that code agents outperform standalone LLMs significantly, with the gain mainly coming from improving correctness. Among standalone LLMs, reasoning models outperform non-reasoning models.}
\label{main results}
\end{table}

\para{Inference} 
We use greedy decoding for all non-reasoning models. For reasoning models, we use their default temperature setting, since they do not expose this parameter. We set a max token response of 3,072 tokens. For OpenAI reasoning models, we set the reasoning effort to medium, and for other reasoning models, we set an additional 8,000 thinking tokens.

\para{Prompt} We design a neutral prompt for all evaluations, which does not provide any security-specific instructions but simply asks the model to fill in the masked region based on the provided context. Unlike the oracle security reminder in \SecCodePLT \cite{yang2024seccodeplt} and \BaxBench \cite{vero2025baxbench} which provide CWE information to avoid task-specific vulnerabilities, our prompt reflects realistic code completion scenarios where developers do not know beforehand what vulnerabilities might be introduced. Therefore, the models must identify and prevent vulnerabilities only by understanding the task.

\subsection{Results}
\para{Overall Performance}
We conduct extensive evaluation on different LLMs and agent setups. Table \ref{main results} presents all results. Among standalone models, \gptFive achieves the highest \secpassone of 39.3\% followed by \oThree at 32.4\%, while the best model of the Claude series achieves 31.1\%, still 8.2\% behind the top OpenAI model. The best Gemini model achieves an even lower score of 27.7\%. The lowest performing model, \llamaThreeEightB, obtained a \secpassone of 5\%, as it struggled with both correctness and security. For agents, \openhands paired with \oThree achieves the best performance of a 53.5\% \secpassone among all setups, which shows a substantial improvement over standalone model evaluation. However, even the top setup fails to generate both secure and correct code for nearly half of the tasks, demonstrating that our repository-level benchmark is quite challenging for both standalone models and agents.

\begin{figure}[t!]
\captionsetup{skip=5pt}
    \centering
    \includegraphics[width=0.47\textwidth]{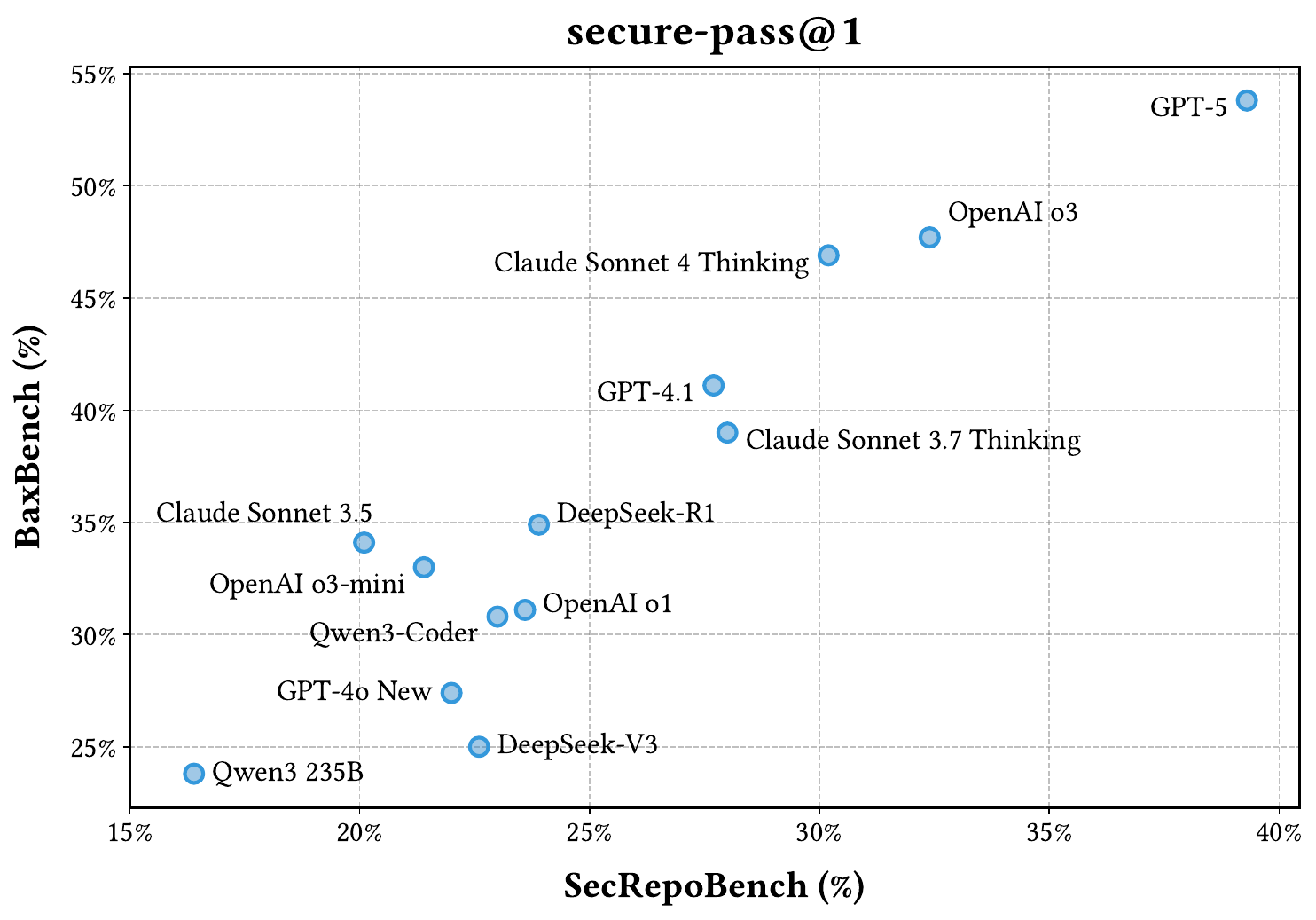}
    \caption{Comparison of \secpassone performance on \BaxBench versus \benchmark. While standalone models achieve relatively high \secpassone on \BaxBench for generating self-contained programs, their performance significantly declines on the more challenging repository-level benchmark \benchmark{}.  Notably, model rankings vary across the two benchmarks, demonstrating that performance on \BaxBench does not generalize well to repository-level code generation tasks in \benchmark. }
    \label{fig:result-highlight}
    \Description{}
\end{figure}

Figure \ref{fig:result-highlight} compares the \secpassone of various top standalone models against \BaxBench \cite{vero2025baxbench}. We observe that the performance of LLMs to generate secure and correct self-contained programs does not generalize to their performance in real-world C/C++ projects. All evaluated models show significantly lower \secpassone on \benchmark compared to \BaxBench. The best model \gptFive has 53.8\% \secpassone on \BaxBench, but only 39.3\% on \benchmark. Notably, even our best agent setup (\openhands + \oThree at 53.5\%) does not surpass the standalone model performance on BaxBench (53.8\%). Meanwhile, we observe that relative model ranking changes when evaluated over the two benchmarks. For example, \claudeSonnetThreeFive and \oThreeMini present better results on \BaxBench than \gptFourONew and \deepseekVThree, but worse on \benchmark. These findings indicate that our benchmark \benchmark is more difficult than \BaxBench.
\begin{table*}[t!]
\centering
\small
\begin{tabular}{@{}l
                    c
                      S[table-format=2.1, table-number-alignment=center]
                      S[table-format=2.1, table-number-alignment=center]
                      S[table-format=2.1, table-number-alignment=center]@{}}
        \toprule
        & & \textbf{secure-pass@1} & \textbf{pass@1} & \textbf{secure \%} \\
        \textbf{Model} & \textbf{Agent Setup}  & \textbf{(\%)} & \textbf{(\%)} & \textbf{(\%)} \\
        \midrule
        \oThree & + \openhands                  & 21.1 & 28.9 & 16.0 \\
        \geminiThreePro & + \openhands          & 20.1 & 21.4 & 28.9 \\
        \claudeSonnetFourFive & + \claudecode   & 19.2 & 22.3 & 20.4 \\
        \geminiThreePro & + \aider              & 17.9 & 22.0 & 22.9 \\
        \oFourMini & + \aider                   & 14.5 & 21.4 & 17.0 \\
        \claudeSonnetFourFive & + \aider        & 14.5 & 4.4 & 19.2 \\
        \claudeSonnetFour & + \aider            & 13.2 & 16.4 & 10.7 \\
        \oFourMini & + \openhands               & 12.6 & 22.3 & 15.4 \\
        \claudeSonnetFourFive & + \openhands    & 12.6 & 21.7 & 14.2 \\
        \claudeSonnetFour & + \openhands        & 12.6 & 17.9 & 16.4 \\
        \gptFive & + \openhands                 & 12.0 & 25.5 & 4.7 \\
        \gptFourOne & + \aider                  & 11.3 & 15.7 & 10.7 \\
        \gptFive & + \aider                     & 11.3 & 15.4 & 4.7 \\
        \oThree & + \aider                      & 11.0 & 13.2 & 9.7 \\
        \gptFourOne & + \openhands              & 1.6 & 6.6 & 7.2 \\
        
        \midrule
        \textbf{Average}        &  & \textbf{13.7} & \textbf{18.3} & \textbf{14.5} \\
        \bottomrule
    \end{tabular}
    \vspace{5pt}
\caption{Percentage improvements of different agent setups on \benchmark{} compared to standalone LLMs. On average, the increase in \passone (18.3\%) is significantly higher than the increase of secure percentage (14.5\%). This suggests that, compared to standalone LLMs, agents focus more on improving the functional correctness of code, but less on the underlying security issues.}
\label{improvement}
\end{table*}

\para{Comparison between Different Setups}

\parai{Code Agents vs. Standalone Models} Table \ref{improvement} lists the performance improvements across agent setups over standalone models. In all cases, agent combinations achieve higher performance compared to their standalone counterparts. Agents could interactively explore the codebase and use tools to refine their code, rather than fixed content provided by the context retriever. Thus, agent integration provides reasonable benefits for repository-level code completion.  

However, the results also reveal a key limitation of the agents, where the improvements in \passone on average are significantly higher than the improvements in secure percentage. For example, \openhands + \gptFive achieves a 25.5\% improvement in correctness, but only has a 4.7\% secure percentage increase. The agent appears to prioritize functional correctness during refinement process but lacks a reliable mechanism to identify and address security vulnerabilities it introduces during code completion, which could lead to severe security risks in real-world deployments.

In contrast, we also observe four outliers. For \aider + \claudeSonnetFourFive, it only adds 4.4\% to its correctness, but improves 19.2\% of the secure percentage. By manually inspecting the conversation trajectories, we found the model did check some of the vulnerabilities (\eg, initialize memory, adjust the array size, check for out-of-bound read). But it might be overly conservative and harm the functional correctness by introducing unnecessary checks or modifications, since 21\% of code generated by \aider + \claudeSonnetFourFive cannot even compile. For \openhands + \gptFourOne, the opposite originates from \gptFourOne struggling to effectively utilize the available tools in the OpenHands framework. Especially for the tool \code{str\_replace\_editor}, the model fails to use it correctly, ending up in dead loops that eventually terminate the interaction. This degrades agent performance below the baseline from some standalone models. For \geminiThreePro, both agent combinations have a smaller improvement in correctness. This is mainly because the standalone \geminiThreePro baseline has a much lower secure percentage comparing to other models in the same tier.

\parai{Reasoning vs. Non-reasoning Models}
Based on Table \ref{main results}, reasoning models perform better compared to non-reasoning models in general. Among standalone LLMs, reasoning models occupy 9 out of the top 10 positions. The extended thinking process allows the models to better understand repository-level context and identify issues before generating code. But this is not always the case. For example, comparing non-reasoning \qwenThreeCoder and reasoning model \qwenThree from the same model family, the former outperforms the latter by 6.6\%. The reasoning models \gptOss and \oThreeMini both provide modest results, surpassed by other more advanced models. This suggests that the quality of the reasoning mechanism and the degree to which the model is calibrated for secure coding tasks also matter.

\subsection{Failure Modes Analysis}
In this section, we examine the details of the best-performing setup \openhands + \oThree. 

\para{Compilation Issues}
Though the code agent is allowed to check the definition of functions and variables by examining the raw code, it still gives 27 tasks (8.5\%) of compilation errors. The most common issue is undeclared identifiers, where the agent attempts to hallucinate reasonable-sounding but non-existent variables, \code{struct} members or macros based on an incomplete understanding of the repository. The second most frequent issue is undefined references, where the agents reference functions not properly linked during compilation, indicating difficulties in understanding cross-file dependencies. Other notable compilation errors include redefinition of existing symbols, type mismatches and conversion errors. Those errors reveal that agents still struggle with maintaining consistency and avoiding hallucination of non-existent code elements.

\para{Correct but Insecure}
We also analyze the tasks that are functionally correct but fail the security test. The most common type is missing conditional-check, where the generated code lacks necessary security-relevant checks in the secure patch. For example, in task 52317 (hunspell), the developer-written patch checks whether the index is less than the size of a word before indexing into it. While the standalone \oThree lacks this check, the agent does not add it either despite having access to the indexing logic. The second common pattern is incorrect conditional check, where the agent attempts to add the security-relevant check but implements it incorrectly. This indicates that while the agent has some awareness of potential security issues, it is unable to fully understand the code logic and implement proper validations. Additional patterns include incorrect memory allocation (\eg, the agent uses the wrong allocation function or miscalculates buffer sizes that lead to an overflow issue), uninitialized memory manipulation and unsafe pointer dereference. This further proves our previous conclusion where the agent seems to prioritize functional correctness over security.

\para{Secure but Incorrect}
While the agent improves a lot in functional correctness, we still observe a few tasks that fail to implement the correct functionality. While this does not give us further information, it shows the effectiveness of our developer-written unit tests, which agrees with previous research \cite{fu2024constrained} that security checks alone are not enough to test the quality of a code completion.

\section{Related Work}

\para{Secure Code Generation Benchmarks} 
Many benchmarks~\cite{chen2021evaluating,liu2024your,austin2021program,hendrycks2measuring,li2022competition,gu2024cruxeval,white2024livebench,jain2024livecodebench,xia2024top,ding2023crosscodeeval,liurepobench,bogomolov2024long,liang2024can} evaluate LLMs on generating correct code. A few benchmarks evaluate how LLMs generate correct code at the repository level~\cite{ding2023crosscodeeval,liurepobench,bogomolov2024long,liang2024can}, but they do not evaluate the security of generated code. Earlier secure code generation benchmarks only evaluate the security of generated code, such as the Copilot dataset~\cite{pearce2022asleep}, \textsc{SecurityEval}~\cite{siddiq2022securityeval}, \textsc{CodeLMSec}~\cite{hajipour2024codelmsec}, \textsc{SafeCoder}~\cite{he2024instruction}, and \textsc{CyberSecEval}~\cite{bhatt2023purple}. Later, researchers developed benchmarks to evaluate both the security and the correctness of generated code~\cite{fu2024constrained,peng2025cweval,yang2024seccodeplt,vero2025baxbench}, since it is important to evaluate both criteria at the same time. \textsc{CWEval}~\cite{peng2025cweval} has demonstrated that it is imprecise to use static analyzer to evaluate the security of generated code. Therefore, recent benchmarks~\cite{yang2024seccodeplt,peng2025cweval,vero2025baxbench} opt for using dynamic security test cases. Concurrent work \textsc{SecCodePLT}~\cite{yang2024seccodeplt} has C/C++ code completion tasks in real-world repositories. One of the main differences between the C/C++ tasks of \textsc{SecCodePLT} and our work is the method to evaluate the correctness of completed code. \textsc{SecCodePLT} uses mutated fuzzing inputs as the correctness test suites. On the contrary, we use developer-written unit tests to evaluate the correctness of completed code. Compared to \textsc{SecCodePLT}, our method is more realistic.

\para{Techniques to Generate Secure Code} Researchers have proposed to use prompt engineering~\cite{homoliak2024enhancing}, prefix tuning~\cite{he2023large}, instruction tuning~\cite{he2024instruction}, specialized decoding~\cite{fu2024constrained,li2024cosec}, specialized prompt optimization~\cite{nazzal2024promsec}, and vulnerability repair~\cite{pearce2023examining,islam2024code,nazzal2024promsec} to make LLMs more likely to generate secure code. In this paper, we are the first to study the secure coding abilities of code agent frameworks in combination with LLMs. We will leave other secure code generation techniques as future work.

\section{Discussions}

\para{Benchmark Coverage}
\benchmark{} covers a subset of distinct OSS-Fuzz crash types from ARVO. In particular, 20 crash types in ARVO that have more than 10 samples, and \benchmark{} covers 17 of them. \benchmark{} does not include every sample from ARVO due to the extensive procedure to select vulnerability samples that satisfy all these criteria: the patch only modifies one function, the function also has valid developer-written unit tests, and we can reproduce the crash by modifying the masked region to vulnerable code. \benchmark{} contains code generation tasks that complete a function, which is the standard way of evaluating the code completion capabilities of LLMs. Vulnerabilities patches that change multiple functions are also useful for constructing other security-relevant tasks in code editing and agentic coding scenarios, which we leave for future work.

\para{Generalization of Test Cases}
We reuse the test cases that work on ground truth vulnerable code for evaluating LLM-generated code, including developer-written unit tests and security test cases found by OSS-Fuzz. We assume that the test cases can be generalized to unseen LLM-generated code. This assumption is also made by state-of-the-art benchmarks~\cite{yang2024seccodeplt,vero2025baxbench} that contain tasks to generate self-contained programs. Moreover, our procedure to evaluate the security of LLM-generated code follows the workflow of OSS-Fuzz to run known security test cases.

\section{Conclusion}

This paper has presented \benchmark{}, a repository-level code completion benchmark, to evaluate code agents for correct and secure code completion abilities in real-world coding scenarios. Our experimental results suggest that \benchmark{} is currently the most challenging benchmark that contains both functional correctness and security tests for evaluating secure code generation. We hope \benchmark{} will contribute to advancing research in secure and correct code generation.

\begin{acks}
    We are grateful to Ethan Baker and Yanjun Fu for exploring the ARVO dataset and language model inference code. We are grateful to Jiacheng Li for his advice on the unit test experiment. This research was supported in part by, Open Philanthropy, an NSF CAREER Award CNS-2442719, generous gifts from Google DeepMind and OpenAI, and the Center for AI Safety Compute Cluster. Any opinions, findings, and conclusions or recommendations expressed in this material are those of the author(s) and do not necessarily reflect the views of the sponsors.
\end{acks}

\bibliographystyle{ACM-Reference-Format}
\bibliography{ref}

@misc{Anthropic2024,
  author = {Anthropic},
  title = {Claude Code},
  year = {2025},
  url = {https://www.claude.com/product/claude-code}
}

@misc{openhands,
      title={{OpenHands: An Open Platform for AI Software Developers as Generalist Agents}},
      author={Xingyao Wang and Boxuan Li and Yufan Song and Frank F. Xu and Xiangru Tang and Mingchen Zhuge and Jiayi Pan and Yueqi Song and Bowen Li and Jaskirat Singh and Hoang H. Tran and Fuqiang Li and Ren Ma and Mingzhang Zheng and Bill Qian and Yanjun Shao and Niklas Muennighoff and Yizhe Zhang and Binyuan Hui and Junyang Lin and Robert Brennan and Hao Peng and Heng Ji and Graham Neubig},
      year={2024},
      eprint={2407.16741},
      archivePrefix={arXiv},
      primaryClass={cs.SE},
      url={https://arxiv.org/abs/2407.16741},
}

@misc{aider2023,
  author       = {Paul Gauthier},
  title        = {Aider: AI-assisted coding in your terminal with GPT},
  year         = {2023},
  howpublished = {\url{https://aider.chat/}},
  note         = {Accessed: 2025-05-15}
}

@inproceedings{shen2021localizing,
author = {Shen, Shiqi and Kolluri, Aashish and Dong, Zhen and Saxena, Prateek and Roychoudhury, Abhik},
title = {Localizing Vulnerabilities Statistically From One Exploit},
year = {2021},
isbn = {9781450382878},
publisher = {Association for Computing Machinery},
address = {New York, NY, USA},
url = {https://doi.org/10.1145/3433210.3437528},
doi = {10.1145/3433210.3437528},
booktitle = {Proceedings of the 2021 ACM Asia Conference on Computer and Communications Security},
pages = {537–549},
numpages = {13},
location = {Virtual Event, Hong Kong},
series = {ASIA CCS '21}
}

@inproceedings{pearce2022asleep,
  title={Asleep at the keyboard? assessing the security of github copilot’s code contributions},
  author={Pearce, Hammond and Ahmad, Baleegh and Tan, Benjamin and Dolan-Gavitt, Brendan and Karri, Ramesh},
  booktitle={2022 IEEE Symposium on Security and Privacy (SP)},
  pages={754--768},
  year={2022},
  organization={IEEE}
}

@inproceedings{siddiq2022securityeval,
  title={SecurityEval dataset: mining vulnerability examples to evaluate machine learning-based code generation techniques},
  author={Siddiq, Mohammed Latif and Santos, Joanna CS},
  booktitle={Proceedings of the 1st International Workshop on Mining Software Repositories Applications for Privacy and Security},
  pages={29--33},
  year={2022}
}

@inproceedings{chen2023diversevul,
  title={Diversevul: A new vulnerable source code dataset for deep learning based vulnerability detection},
  author={Chen, Yizheng and Ding, Zhoujie and Alowain, Lamya and Chen, Xinyun and Wagner, David},
  booktitle={Proceedings of the 26th International Symposium on Research in Attacks, Intrusions and Defenses},
  pages={654--668},
  year={2023}
}

@misc{copilot-users-2025,
  author = {{Business of Apps}},
  title = {{Microsoft Copilot Revenue and Usage Statistics (2025)}},
  howpublished = "\url{https://www.businessofapps.com/data/microsoft-copilot-statistics/}",
  year= {2025},
}

@misc{copilot-productivity,
  author = {{Eirini Kalliamvakou, GitHub Blog}},
  title = {{Research: quantifying GitHub Copilot’s impact on developer productivity and happiness}},
  howpublished = "\url{https://github.blog/2022-09-07-research-quantifying-github-copilots-impact-on-developer-productivity-and-happiness/}",
  year= {2022},
}

@misc{code-completion-productivity,
  author = {{Maxim Tabachnyk and Stoyan Nikolov, Google Research}},
  title = {{ML-Enhanced Code Completion Improves Developer Productivity}},
  howpublished = "\url{https://research.google/blog/ml-enhanced-code-completion-improves-developer-productivity/}",
  year= {2022},
}

@inproceedings{he2023large,
  title={{Large language models for code: Security hardening and adversarial testing}},
  author={He, Jingxuan and Vechev, Martin},
  booktitle={{Proceedings of the 2023 ACM SIGSAC Conference on Computer and Communications Security}},
  pages={1865--1879},
  year={2023}
}

@inproceedings{hajipour2024codelmsec,
  title={{CodeLMSec Benchmark: Systematically Evaluating and Finding Security Vulnerabilities in Black-Box Code Language Models}},
  author={Hajipour, Hossein and Hassler, Keno and Holz, Thorsten and Sch{\"o}nherr, Lea and Fritz, Mario},
  booktitle={2024 IEEE Conference on Secure and Trustworthy Machine Learning (SaTML)},
  pages={684--709},
  year={2024},
  organization={IEEE}
}

@inproceedings{he2024instruction,
  title={Instruction Tuning for Secure Code Generation},
  author={He, Jingxuan and Vero, Mark and Krasnopolska, Gabriela and Vechev, Martin},
  booktitle={Proceedings of the International Conference on Machine Learning (ICML)},
  year={2024}
}

@article{bhatt2023purple,
  title={{Purple llama cyberseceval: A secure coding benchmark for language models}},
  author={Bhatt, Manish and Chennabasappa, Sahana and Nikolaidis, Cyrus and Wan, Shengye and Evtimov, Ivan and Gabi, Dominik and Song, Daniel and Ahmad, Faizan and Aschermann, Cornelius and Fontana, Lorenzo and others},
  journal={arXiv preprint arXiv:2312.04724},
  year={2023}
}

@article{homoliak2024enhancing,
  title={{Enhancing Security of AI-Based Code Synthesis with GitHub Copilot via Cheap and Efficient Prompt-Engineering}},
  author={Homoliak, Ivan and Pere{\v{s}}{\'\i}ni, Martin and Smr{\v{c}}ka, Ale{\v{s}} and Malinka, Kamil and Hanacek, Petr},
  journal={arXiv preprint arXiv:2403.12671},
  year={2024}
}

@article{austin2021program,
  title={{Program Synthesis with Large Language Models}},
  author={Austin, Jacob and Odena, Augustus and Nye, Maxwell and Bosma, Maarten and Michalewski, Henryk and Dohan, David and Jiang, Ellen and Cai, Carrie and Terry, Michael and Le, Quoc and others},
  journal={arXiv preprint arXiv:2108.07732},
  year={2021}
}

@article{chen2021evaluating,
  title={Evaluating large language models trained on code},
  author={Chen, Mark and Tworek, Jerry and Jun, Heewoo and Yuan, Qiming and Pinto, Henrique Ponde de Oliveira and Kaplan, Jared and Edwards, Harri and Burda, Yuri and Joseph, Nicholas and Brockman, Greg and others},
  journal={arXiv preprint arXiv:2107.03374},
  year={2021}
}

@inproceedings{pearce2023examining,
  title={{Examining Zero-Shot Vulnerability Repair with Large Language Models}},
  author={Pearce, Hammond and Tan, Benjamin and Ahmad, Baleegh and Karri, Ramesh and Dolan-Gavitt, Brendan},
  booktitle={2023 IEEE Symposium on Security and Privacy (SP)},
  pages={2339--2356},
  year={2023},
  organization={IEEE}
}

@inproceedings{islam2024code,
  title={{Code Security Vulnerability Repair Using Reinforcement Learning with Large Language Models}},
  author={Islam, Nafis Tanveer and Najafirad, Peyman},
  booktitle={Proceedings of the AAAI Conference on Artificial Intelligence Workshop},
  year={2024}
}

@inproceedings{ding2023crosscodeeval,
  title={CrossCodeEval: A Diverse and Multilingual Benchmark for Cross-File Code Completion},
  author={Yangruibo Ding and Zijian Wang and Wasi Uddin Ahmad and Hantian Ding and Ming Tan and Nihal Jain and Murali Krishna Ramanathan and Ramesh Nallapati and Parminder Bhatia and Dan Roth and Bing Xiang},
  booktitle={Thirty-seventh Conference on Neural Information Processing Systems Datasets and Benchmarks Track},
  year={2023},
  url={https://openreview.net/forum?id=wgDcbBMSfh}
}

@misc{mei2024arvo,
      title={ARVO: Atlas of Reproducible Vulnerabilities for Open Source Software}, 
      author={Xiang Mei and Pulkit Singh Singaria and Jordi Del Castillo and Haoran Xi and Abdelouahab and Benchikh and Tiffany Bao and Ruoyu Wang and Yan Shoshitaishvili and Adam Doupé and Hammond Pearce and Brendan Dolan-Gavitt},
      year={2024},
      journal={arXiv preprint arXiv:2408.02153}, 
}

@inproceedings{nazzal2024promsec,
  title={{PromSec: Prompt Optimization for Secure Generation of Functional Source Code with Large Language Models (LLMs)}},
  author={Nazzal, Mahmoud and Khalil, Issa and Khreishah, Abdallah and Phan, NhatHai},
  booktitle={Proceedings of the 2024 on ACM SIGSAC Conference on Computer and Communications Security},
  pages={2266--2280},
  year={2024}
}

@article{wang2023enhancing,
  title={Enhancing Large Language Models for Secure Code Generation: A Dataset-driven Study on Vulnerability Mitigation},
  author={Wang, Jiexin and Cao, Liuwen and Luo, Xitong and Zhou, Zhiping and Xie, Jiayuan and Jatowt, Adam and Cai, Yi},
  journal={arXiv preprint arXiv:2310.16263},
  year={2023}
}

@inproceedings{yang2024seccodeplt,
  title={SecCodePLT: A Unified Benchmark for Evaluating the Security Risks and Capabilities of Code Agents},
  author={Nie, Yuzhou and Wang, Zhun and Yang, Yu and Jiang, Ruizhe and Tang, Yuheng and Davies, Xander and Gal, Yarin and Li, Bo and Guo, Wenbo and Song, Dawn},
  booktitle={The Thirty-Ninth Annual Conference on Neural Information Processing Systems},
  year={2025}
}

@inproceedings{zhong2024can,
  title={Can LLM Replace Stack Overflow? A Study on Robustness and Reliability of Large Language Model Code Generation},
  author={Zhong, Li and Wang, Zilong},
  booktitle={Proceedings of the AAAI Conference on Artificial Intelligence},
  year={2024}
}

@inproceedings{hendrycks2measuring,
  title={Measuring Coding Challenge Competence With APPS},
  author={Hendrycks, Dan and Basart, Steven and Kadavath, Saurav and Mazeika, Mantas and Arora, Akul and Guo, Ethan and Burns, Collin and Puranik, Samir and He, Horace and Song, Dawn and others},
  booktitle={Thirty-fifth Conference on Neural Information Processing Systems Datasets and Benchmarks Track (Round 2)},
  year={2021}
}

@article{li2022competition,
  title={Competition-level code generation with alphacode},
  author={Li, Yujia and Choi, David and Chung, Junyoung and Kushman, Nate and Schrittwieser, Julian and Leblond, R{\'e}mi and Eccles, Tom and Keeling, James and Gimeno, Felix and Dal Lago, Agustin and others},
  journal={Science},
  volume={378},
  number={6624},
  pages={1092--1097},
  year={2022},
  publisher={American Association for the Advancement of Science}
}

@article{liu2024your,
  title={Is your code generated by chatgpt really correct? rigorous evaluation of large language models for code generation},
  author={Liu, Jiawei and Xia, Chunqiu Steven and Wang, Yuyao and Zhang, Lingming},
  journal={Advances in Neural Information Processing Systems},
  volume={36},
  year={2024}
}

@article{xia2024top,
  title={Top Leaderboard Ranking= Top Coding Proficiency, Always? EvoEval: Evolving Coding Benchmarks via LLM},
  author={Xia, Chunqiu Steven and Deng, Yinlin and Zhang, Lingming},
  journal={Proceedings of the First Conference on Language Modeling},
  year={2024}
}

@article{gu2024cruxeval,
  title={Cruxeval: A benchmark for code reasoning, understanding and execution},
  author={Gu, Alex and Rozi{\`e}re, Baptiste and Leather, Hugh and Solar-Lezama, Armando and Synnaeve, Gabriel and Wang, Sida I},
  journal={arXiv preprint arXiv:2401.03065},
  year={2024}
}

@article{white2024livebench,
  title={Livebench: A challenging, contamination-free llm benchmark},
  author={White, Colin and Dooley, Samuel and Roberts, Manley and Pal, Arka and Feuer, Ben and Jain, Siddhartha and Shwartz-Ziv, Ravid and Jain, Neel and Saifullah, Khalid and Naidu, Siddartha and others},
  journal={arXiv preprint arXiv:2406.19314},
  year={2024}
}

@article{jain2024livecodebench,
    title={LiveCodeBench: Holistic and Contamination Free Evaluation of Large Language Models for Code},
    author={Jain, Naman and Han, King and Gu, Alex and Li, Wen-Ding and Yan, Fanjia and Zhang, Tianjun and Wang, Sida and Solar-Lezama, Armando and Sen, Koushik and Stoica, Ion},
    journal={arXiv preprint arXiv:2403.07974},
    year={2024}
}

@inproceedings{liurepobench,
  title={RepoBench: Benchmarking Repository-Level Code Auto-Completion Systems},
  author={Liu, Tianyang and Xu, Canwen and McAuley, Julian},
  booktitle={The Twelfth International Conference on Learning Representations},
  year={2024}
}

@article{bogomolov2024long,
  title={Long Code Arena: a Set of Benchmarks for Long-Context Code Models},
  author={Bogomolov, Egor and Eliseeva, Aleksandra and Galimzyanov, Timur and Glukhov, Evgeniy and Shapkin, Anton and Tigina, Maria and Golubev, Yaroslav and Kovrigin, Alexander and van Deursen, Arie and Izadi, Maliheh and others},
  journal={arXiv preprint arXiv:2406.11612},
  year={2024}
}

@article{liang2024can,
  title={Can Language Models Replace Programmers? REPOCOD Says' Not Yet'},
  author={Liang, Shanchao and Hu, Yiran and Jiang, Nan and Tan, Lin},
  journal={arXiv preprint arXiv:2410.21647},
  year={2024}
}

@article{fu2024constrained,
  title={{Constrained Decoding for Secure Code Generation}},
  author={Yanjun Fu and Ethan Baker and Ding, Yu and Chen, Yizheng},
  journal={arXiv:2405.00218},
  year={2024}
}

@article{vero2025baxbench,
  title={BaxBench: Can LLMs Generate Correct and Secure Backends?},
  author={Vero, Mark and M{\"u}ndler, Niels and Chibotaru, Victor and Raychev, Veselin and Baader, Maximilian and Jovanovi{\'c}, Nikola and He, Jingxuan and Vechev, Martin},
  journal={arXiv preprint arXiv:2502.11844},
  year={2025}
}

@inproceedings{peng2025cweval,
  title={CWEval: Outcome-driven Evaluation on Functionality and Security of LLM Code Generation},
  author={Peng, Jinjun and Cui, Leyi and Huang, Kele and Yang, Junfeng and Ray, Baishakhi},
  booktitle={Proceedings of the Second International Workshop on Large Language Models for Code},
  year={2025}
}

@inproceedings{li2024cosec,
  title={CoSec: On-the-Fly Security Hardening of Code LLMs via Supervised Co-decoding},
  author={Li, Dong and Yan, Meng and Zhang, Yaosheng and Liu, Zhongxin and Liu, Chao and Zhang, Xiaohong and Chen, Ting and Lo, David},
  booktitle={Proceedings of the 33rd ACM SIGSOFT International Symposium on Software Testing and Analysis},
  pages={1428--1439},
  year={2024}
}

@inproceedings{ding2025vulnerability,
  title={Vulnerability detection with code language models: How far are we?},
  author={Ding, Yangruibo and Fu, Yanjun and Ibrahim, Omniyyah and Sitawarin, Chawin and Chen, Xinyun and Alomair, Basel and Wagner, David and Ray, Baishakhi and Chen, Yizheng},
  booktitle={IEEE/ACM International Conference on Software Engineering \textbf{(ICSE)}},
  year={2025}
}

\end{document}